# Post-Growth Shaping and Transport Anisotropy in 2D InAs Nanofins

J. Seidl,[1] J.G. Gluschke,[1] X. Yuan[2,†], H.H. Tan[2], C. Jagadish[2], P. Caroff[2,§] and A.P. Micolich[1,*]

[1]School of Physics, University of New South Wales, Sydney NSW 2052, Australia

[2]ARC Centre of Excellence for Transformative Meta-Optical Systems, Department of Electronic Materials Engineering, Research School of Physics, The Australian National University, Canberra ACT 2601, Australia

**ABSTRACT**

We report on the post-growth shaping of free-standing 2D InAs nanofins that are grown by selective-area epitaxy and mechanically transferred to a separate substrate for device fabrication. We use a citric acid based wet etch that enables complex shapes, *e.g.*, van der Pauw cloverleaf structures, with patterning resolution down to 150 nm as well as partial thinning of the nanofin to improve the gate response. We exploit the high sensitivity of the cloverleaf structures to transport anisotropy to address the fundamental question of whether there is a measurable transport anisotropy arising from wurtzite/zincblende polytypism in 2D InAs nanostructures. We demonstrate a mobility anisotropy of order 2-4 at room temperature arising from polytypic stacking faults in our nanofins. Our work highlights a key materials consideration for devices featuring self-assembled 2D III-V nanostructures using advanced epitaxy methods.

**Current Addresses**

§ Microsoft Quantum Lab Delft, Delft University of Technology, 2600 GA Delft, The Netherlands.

† Hunan Key Laboratory of Nanophotonics and Devices, School of Physics and Electronics, Central South University, 932 South Lushan Road, Changsha, Hunan 410083, P. R. China.

**INTRODUCTION**

The challenges of integrating III-V semiconductors on Si[1] and developing Majorana-based quantum computation structures[2] have been strong drivers for innovation in the self-assembled growth of III-V nanostructures beyond the basic 1D vertical nanowire morphology.[3] A common early approach was to use a vapor-liquid-solid (VLS) grown nanowire as a template to make branches,[4] intersections[5,6] (nanocrosses) and sail/fin structures.[7–12] 2D nanostructures have also been grown, direct from the Au catalyst for VLS growth,[13] as well as from nanoscale apertures in a $SiO_2$ layer on a crystalline substrate.[14] The latter is an example of a technique known as selective-area epitaxy (SAE) pioneered in the 1960s whereby semiconductor crystal growth is templated using lithographically-patterned windows etched into an amorphous oxide layer.[15,16] Several decades followed before growth and nanoscale patterning methods reached sufficient maturity to enable high-uniformity, large-area nanowire arrays,[17,18] with the small interfacial area facilitating effective III-V integration on Si.[19] More recently, the two drivers mentioned earlier have produced an explosion of nanostructures grown by SAE including horizontal 1D In(Ga)As nanowires,[20–22] nanocrosses[23] and nanowire crosshatch networks,[22,24] as well as taller nanofin structures composed of InAs,[25,26] GaN,[27] InP,[28,29] GaAs[30,31] and GaAs with an InAs strip on top.[32] The ability to produce high-uniformity arrays with substantial scope for variations in shape and composition offers selective-area epitaxy an exciting future.[33] In this paper, we address two challenges for 2D nanofin structures grown by SAE: post-growth shape control and the possibility of transport anisotropy due to crystallographic polytypism.

Regarding post-growth shape control, a common theme is the quest to increase the geometrical diversity of self-assembled semiconductor nanostructures beyond the basic 1D nanowire morphology that has dominated the past twenty or so years.[22,24,29,33–35] We recently reported on free-standing rectangular InAs nanofins grown by SAE where the height, width and thickness can be tuned *via* the template and growth conditions.[26] These nanofins can be detached from the growth substrate and transferred using a micromanipulator to sit flat on a separate prepatterned device substrate for electrical characterization.[26,36] This offers significant versatility in choice of device substrate and the ability to position the nanofin on top of pre-existing gate structures, as demonstrated for InAs nanowires.[37,38] The rectangular 2D nanofin morphology offers an ideal foundation for more complex geometries. Here we show this can be achieved using electron-beam lithography and wet etching after transferring the nanofins to a device substrate. We obtain pattern resolution down to 150 nm with sufficient etch rate control to either etch completely through the nanofin to impose a shape or partially thin the nanofin to locally increase the gate response. This type of local thickness modulation cannot be directly achieved using any of the growth methods discussed earlier and adds an additional degree of freedom for device design.

Turning to transport properties, a key feature of III-V nanowires is polytypism, *i.e.*, the ability to stably exhibit segments with wurtzite (WZ) and/or zincblende (ZB) crystal structure, by virtue of their tiny cross-sectional area.[39,40] Transitions between crystal phase, *i.e.*, stacking faults, occur perpendicular to the nanowire growth axis, and are randomly distributed or can be controlled, depending on the growth conditions.[41,42] Polytypism in nanowires is known to significantly increase electrical resistivity[43,44] and can be put to practical use, *e.g.*, in making quantum dots,[45–47] or eliminated entirely to optimize electronic performance.[43,48] The high aspect ratio of nanowires combined with the difficulties in making electrical contacts to the sides of the nanowire, *e.g.*, for Hall effect measurements,[49,50] means that any transport anisotropy arising from polytypism is difficult to measure. The situation differs markedly for 2D morphologies like nanofins, where the aspect ratio is much smaller and can even be significantly less than 1, *i.e.*, the sample is longer along the stacking fault planes than perpendicular to them, and significant scope for multi-terminal devices with current density components in multiple crystallographic directions exists. Thus an important question in moving beyond nanowires to 2D growth by SAE is: Does polytypism generate transport anisotropy in these 2D III-V nanomaterials, and if so, how large is it?

A widely used approach to accurately measuring electronic properties is the van der Pauw method,[51,52] which involves making four contacts to a flat lamella and cycling through all contact permutations with a current $I$ driven *via* one adjacent contact pair and the other adjacent pair used to measure the resulting voltage $V$, giving a van der Pauw resistance $R_{vdP} = V/I$. The common assumption is that the lamella is homogeneous and isotropic,[51–54] *i.e.*, described entirely by a single valued resistivity $\rho$. However, more recent work has shown that the approach can also be effectively used for semiconductors with anisotropic transport properties.[55] Notably, Bierwagen *et al.*[55] report that the van der Pauw method is significantly more sensitive to anisotropy than, *e.g.*, Hall-bar measurements or two-terminal directional studies, a finding our own data here supports (see Supporting information). A key aspect is that the van der Pauw approach enables any asymmetry arising from the contacts or geometry to be separated from the underlying materials anisotropy. In a sense, the van der Pauw approach can act as an 'anisotropy amplifier' that enables more clear, unambiguous determination of anisotropy, particularly in cases where it might be relatively small.

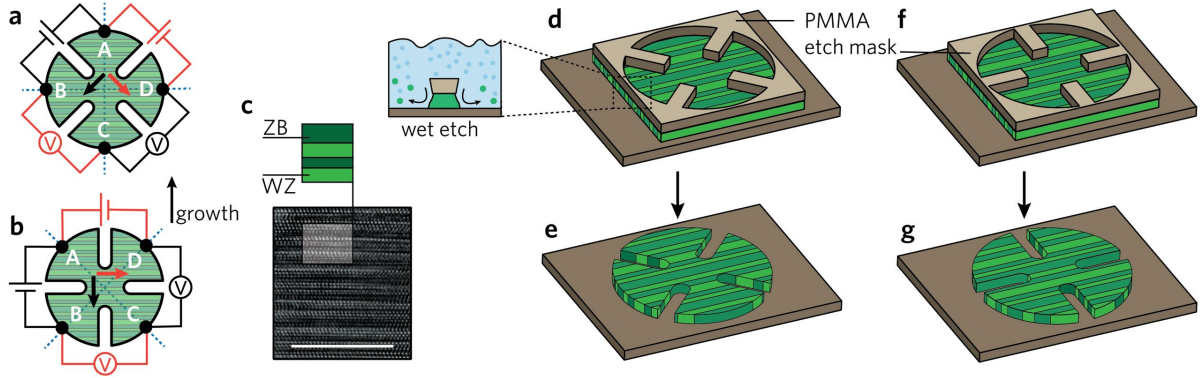

**Figure 1. a/b** Schematic of cloverleaf structures relative to underlying crystallographic polytypism to illustrate the orientation that **a** does not (Orientation 1) and **b** does (Orientation 2) show transport anisotropy in van der Pauw measurements. The wurtzite (WZ) and zincblende (ZB) phases are shaded light and dark green respectively. The cloverleaf axes are indicated by the blue dashed lines. The two distinct subsets of van der Pauw circuit arrangement are indicated in red/black (see main text). The contacts are labelled A-D, as used later in Fig. 3/4. The red and black arrows indicate the main potential difference and current directions, as discussed later in Fig. 3f/g. **c** shows a TEM image for our nanofins. The stripes apparent in the image originate from the polytype stacking faults (see also Fig. 4b). The contrast of the image was increased for clarity. The scale bar represents 10 nm. **d/f** A polymethylmethacrylate (PMMA) mask patterned using electron-beam lithography protects part of the nanofin during the wet etch (see inset to **d**). **e/g** The mask sets the orientation of the final device with respect to the underlying crystal orientation, which for **g**, enables measurement of the transport anisotropy as explained in the text and **b**.

There are two important considerations to our intended application – contacts and orientation. Regarding contacts, the van der Pauw method is generally predicated on the use of electrical contacts that are negligible in size compared to the sample and located on its periphery. This is often difficult to achieve, and essentially impossible for nanoscale structures from a practical standpoint.[36] The solution offered by van der Pauw[52] is the cloverleaf geometry in Fig. 1, which features four large semiconductor 'pads' connected to a small square central measurement region. These recessed pads mitigate the distorting impact of the relatively large metal contacts overlapping the nanostructure, which are needed to obtain reliable low resistance ohmic contacts. We will later show that our 2 μm × 2 μm nanofins can be etched into a cloverleaf structure with reliable low resistance contacts and relatively low background asymmetry, *e.g.*, due to imperfections in contacts/geometry.

We find that the orientation of the van der Pauw geometry relative to the material's anisotropy axes, in this case the growth axis/stacking fault planes, is crucial to effective measurement. This is best understood by reference to Fig. 1a/b, where two orientations relative to the growth axis are shown. Orientation 1 (Fig. 1a) has the cloverleaf axes (blue dashed lines) running parallel and perpendicular to the stacking fault planes; the WZ and ZB phases appear in light/dark green, respectively. Orientation 2 (Fig. 1b) is at 45° with respect to Orientation 1. The anisotropy is determined by comparing the van der Pauw resistance obtained from two orthogonal measurement configurations (red and black). Note that for Orientation 1, the two measurement configurations are symmetrical, with the current running at a 45° angle to the stacking fault planes in both cases (red/black arrows in Fig. 1a). This means both arrangements give nominally identical van der Pauw resistance. In contrast, for Orientation 2, the current runs parallel/perpendicular to the stacking fault planes, producing to a significant difference in the two van der Pauw resistances arising from the underlying transport anisotropy.

To bring this into the context of the state-of-the-art for the field, we note that our cloverleaf structures are topologically equivalent to the InAs cross-junctions studied by Schmid *et al.*[20] (blue dashed lines in Fig. 1a/b are guide to the eye). Notably, their van der Pauw measurements revealed no resistance anisotropy within error despite the presence of stacking faults in the structure.[20] This is surprising at first sight, given the significant effect of stacking faults on transport in nanowires,[43,44] but can be readily explained by reference to Fig. 1. The device in Ref. 20 was grown directly into a cross-shaped template from a Si seed at the end of one arm of the cross and thus effectively resembles Orientation 1. For either van der Pauw configuration in Orientation 1, the current runs at 45° to the stacking fault planes in the central region, masking any anisotropy arising from polytypism in the cross-junction. To see polytypism-driven anisotropy, one instead needs to be in Orientation 2, and this orientation, either as a cloverleaf or a topologically equivalent cross-junction, cannot presently be directly grown by selective-area epitaxy approaches to our knowledge. This highlights an important point of significance in our work. Our ability to grow 2D nanofins, transfer them to a separate substrate, and then arbitrarily orient the cloverleaf structure relative to the growth axis using post-growth patterning (Fig. 1d-g) is crucial to the ability to address the anisotropy arising from polytypism in 2D III-V nanostructures. The approach may well find utility on other anisotropic 2D materials also.

Ultimately, our van der Pauw measurements show that WZ/ZB polytypism generates a measurable transport (conductivity) anisotropy in our nanofins that can be tuned between 2 and 11 based on temperature and carrier density, as controlled by the back-gate voltage. This corresponds to a difference in the underpinning electron mobility of a factor of 2-4 at room temperature depending on orientation.

**RESULTS/DISCUSSION**

The nanofins were grown by metal-organic vapor-phase epitaxy (MOVPE) on an InP (111)B substrate with a prepatterned SiO$_x$ layer to facilitate selective-area epitaxy.[26] We obtain 2D structures with a well-defined footprint that is set by initial nucleation templated by the SiO$_x$ mask and thereafter maintained by the strong preference for this system to favor formation and stability of six {110} vertical side-facets.[26,29] We obtain a flat {111}B top-facet to get a fully rectangular shape by limiting the growth temperature and V/III ratio.[26] Full details for the growth are reported elsewhere.[26] Using this approach we get large arrays of uniform nanofins as shown in the scanning electron microscope (SEM) image in Fig. 2a. We used two different batches of nanofins for our study. The low aspect ratio nanofins are approximately 2 μm high, 2 μm wide, have thickness 55-65 nm and were grown with a V/III ratio of 664 at 675°C for 300 s. The low aspect ratio nanofins were used for the cloverleaf structures in Fig. 3. The high aspect ratio fins are 5 μm high, 1 μm wide, have two different thicknesses ~60 nm and ~150 nm and were grown with a V/III ratio of 1000 at 720°C for 300 s. The thicker high aspect ratio nanofins were used for the etch studies in Fig. 2. The thinner high aspect ratio nanofins were used for the measurements in Fig. 4. We used a custom micromanipulator setup with an ultrasharp needle (0.1 μm tip) to transfer individual InAs nanofins from a vertical orientation on the growth substrate to a horizontal orientation, *i.e.*, laying flat, on a separate device substrate. The nanofins generally fracture cleanly at their base, preserving a clean rectangular shape providing the width does not significantly exceed the height of the nanofin. The micromanipulator transfer enables nanofin placement with micrometer accuracy on the target substrates with relatively little material wastage. We placed the nanofins on a Si/SiO$_2$ device substrate with prepatterned alignment markers and bond pads for the fabrication of multi-terminal devices. The highly doped Si substrate is used as a back-gate to electrostatically tune the carrier density in the nanofins.[26] Further shaping of the nanofins occurs as follows. An etch mask was exposed in polymethylmethacrylate (PMMA) resist (see Fig. 1d/f) using electron-beam lithography (EBL). A short oxygen plasma etch (30-60 s) was used to remove any resist residue. The sample was then submerged in a solution of citric acid and hydrogen peroxide at 25ºC for approximately 30 s to remove the exposed InAs (see Fig. 1d inset). Our etchant stock solution was a 1:20 mixture of 50% aqueous citric acid solution and 30% aqueous $H_2O_2$ solution, which can be diluted further with $H_2O$ to reduce the etch rate. Segments of the nanofin can be either thinned or removed completely in the exposed area depending on the etch time as shown in Fig. 2. The PMMA was not noticeably affected by the etch and did not require hard baking. The PMMA was removed by a 5 min acetone soak at room temperature leaving the etched nanofin structure. The final step was to add contacts to the etched nanofin, which was achieved by an EBL-patterned metallization of 5-6 nm Ni and 125-195 nm Au by thermal evaporation, adjusted to nanofin thickness. To minimize the contact resistance, the EBL-patterned sample underwent a 30 s oxygen plasma etch to remove resist residue and a 120 s passivation treatment in $(NH_4)_2S_x$ solution at 40ºC to remove native oxide and sulfur-passivate the InAs surface immediately prior to loading into the thermal evaporator. Figures 2b/c show SEM images of completed cloverleaf and cross-junction devices, each with four ohmic contacts to demonstrate our shaping capability. Both devices were fabricated from nominally identical 2 μm × 2 μm nanofins. We will return to the cloverleaf devices in Fig. 3.

**Etch characterization:** The ability to obtain etched shape fidelity and controlled thinning of the nanostructure is reliant on an accurate determination of basic etch characteristics such as etch rate. We measure the etch rate using a sequence of EBL-patterned etches, each with a fresh EBL resist and a different etch time $t$. This generates a series of narrow etched strips, nominally 500 nm wide, spanning the nanofin, for which we measure the etch depth $d$ using an AFM. An AFM image of a nanofin with four etched strips with different $t$ is shown in Fig. 2d(inset). Figure 2d shows $d$ versus $t$ for the stock solution and two dilutions (1:2 and 1:5 in $H_2O$). There is a clear linear dependence of $d$ on $t$ in each case, with the stock solution giving an etch rate of $1.9 \pm 0.2$ nm/s with low surface roughness (1.7 nm rms) in the etched regions (see Supplementary Figure S1 for an AFM line scan). The stock solution etch rate is sufficient for controlled thinning with ~10% accuracy in etch depth.

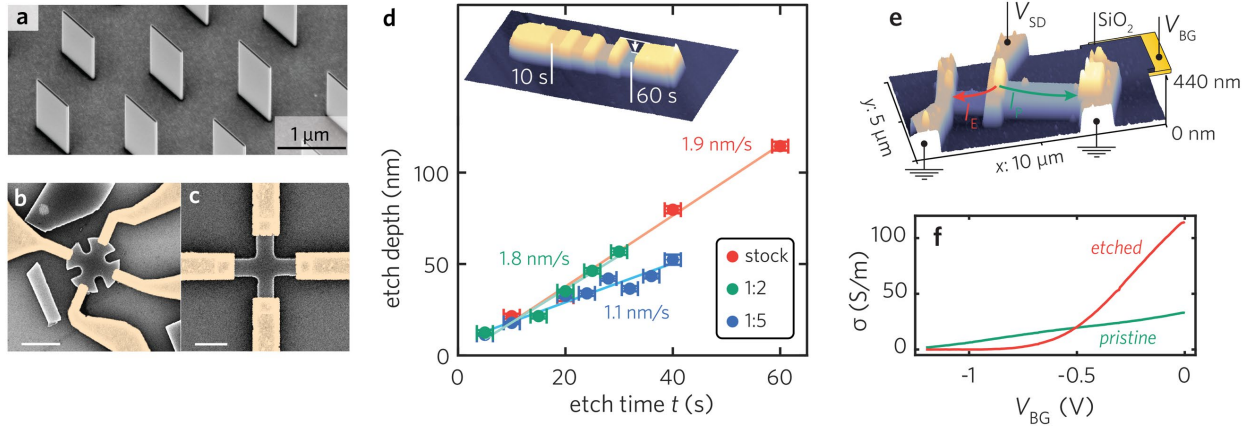

**Figure 2.** False color SEM images of **a** an InAs nanofin array after growth, **b** completed van der Pauw cloverleaf and **c** cross-junction devices. The scale bars in **b/c** are 1 μm. **d** etch depth *vs* time for three different etch dilutions. The uncertainty is one standard deviation for etch depth and 1.5 s for etch time. The inset shows a corresponding AFM image for a high aspect ratio InAs nanofin with four stripes with differing etch time for the stock solution. **e** AFM image of a back-gated nanofin common-source field-effect transistor pair to study the effect on gate response of thinning the nanofin. The transistor on the right is pristine (150 nm thickness) and the transistor on the left has been thinned by 80 nm. **f** Conductivity $\sigma$ *vs* back-gate voltage $V_{BG}$ for the etched (red) and pristine (green) transistors in **e**. Data was obtained at temperature $T = 70$ K.

The dilution profile appears to be non-linear; the etch rate for the 1:2 dilution matches the stock solution within uncertainty while the 1:5 dilution gives a significant reduction in etch rate to $1.1 \pm 0.2$ nm/s, enabling more precise thinning to be obtained, if desired.

Shaped structures such as those in Fig. 2b/c are obtained by setting $t$ such that $d$ exceeds the nanofin thickness by ~10%. The citric acid etchant is isotropic (see Supplementary Figure S2), so longer etch times only degrade shape fidelity due to gradual resist undercut and rounding at sharp corners. Even for the minimal $t$ required for etch completion, *i.e.*, removing all InAs down to the $SiO_x$ substrate surface, there is some rounding of the desired shape. The most notable implication of the etchant isotropy is a limit on the finest features that can be defined. For example, using an etch mask that defines a narrow straight channel on the top surface of the nanofin, the width at the bottom surface will be twice the nanofin thickness, *i.e.*, for a 30 nm wide etch mask and a 60 nm thick nanofin, the resulting channel will be 150 nm wide.

**Electrical characterization of partially thinned nanofin:** Finite-element modelling in our earlier work on these nanofins suggested that the gate characteristics could be tuned by changing the nanofin thickness.[26] This could be achieved by growing thinner nanofins, as we suggested in Ref. 26, or using a thinning etch, as we now demonstrate. Figure 2e shows a transistor pair with a common source made on a single nanofin such that the channel of one of the two transistors is pristine (unetched) nanofin and the other has a nanofin segment that has been thinned by 80 nm using the stock etchant for $t = 50$ s. Electrical measurements were performed at a temperature $T = 70$ K with a voltage $V_{SD} = 300$ µV applied to the central common source contact. The resulting current flowing to ground *via* the pristine channel $I_P$ and the etched channel $I_E$ was measured simultaneously at the two outer electrodes. Figure 2f shows the corresponding conductivity $\sigma = I_P \ell_P / V_{SD} A_P$ or $I_E \ell_E / V_{SD} A_E$ *versus* back-gate voltage $V_{BG}$ for the two channels, where $\ell_P$, $\ell_E$ and $A_P$, $A_E$ are the corresponding channel lengths and cross-sectional areas obtained from AFM measurements. The pristine channel has $\ell_P = 3.2$ µm and $A_P = 1.2$ µm × 150 nm and the etched channel has $\ell_E = 1.2$ µm and $A_E = 1.08$ µm × 70 nm. The pristine channel (green trace) shows a gradual reduction in $\sigma$ as $V_{BG}$ is made more negative, consistent with the gradual depletion of electrons, as expected for InAs. There are two key differences for the etched channel: a higher $\sigma$ at $V_{BG} = 0$ and a much higher transconductance $\partial\sigma/\partial V_{BG}$. We attribute the higher $\sigma$ to an increased contribution of the surface electron accumulation layer. Dangling bonds at the InAs surface pin the Fermi level in the conduction band edge resulting in an electron density that is significantly higher than in the nanofin core.[56,57] The thinning increases the surface-to-volume ratio of the nanofin leading to an overall higher carrier density in the nanofin, which increases $\sigma$. A side-effect is a higher capacitance-to-volume ratio for the nanofin because the back-gate capacitance is proportional to the nanofin area assuming a parallel-plate capacitor model, and the etch reduces the volume of the channel without reducing its area. This should manifest as stronger gating and is consistent with the increased transconductance that we see for the etched channel in Fig. 2f. This behavior is consistent with the predictions of our earlier work.[26] Ultimately, the transport in the etched segment does not appear to be adversely affected. This is in contrast to other etch methods applied to semiconductor structures that are based on plasma etching. Those methods often cause significant damage to the semiconductor due to lattice displacement, projectile implantation and unintended chemical reactions.[58–60] This can lead to increased electron scattering, which degrades the transport properties.[60,61]

**Polytypism and anisotropy:** We now focus on the two device structures in Fig. 3a/c, both of which started as square nanofins (green dashed line), and were etched into cloverleaf structures with two different orientations relative to the growth axis. We denote these as Orientation 1 (Fig. 3a) and Orientation 2 (Fig. 3b), with Orientation 2 rotated 45° counter-clockwise relative to Orientation 1 subject to fixed nanofin orientation. These two orientations directly correspond to those in Fig. 1a and Fig. 1b, respectively. We chose the low aspect ratio (square) nanofins for the cloverleaf devices as they have twice the width of the high aspect ratio (rectangular) nanofins and thereby provide sufficient space for the cloverleaf pattern to fit. A challenge with the square nanofins is that we cannot precisely track the growth axis during positioning on the device substrate. We will discuss how we overcome this issue when we get to Fig. 4, but in Fig. 3 we use the high and low mobility directions as proxies for materials axes, connecting them to crystallographic directions later. The four ohmic contacts are labelled A-D in counter-clockwise order. The four contacts have 24 possible permutations broken into two sets – van der Pauw and Hall. The van der Pauw set have current flowing *via* adjacent contacts with the contacts for sensing voltage also adjacent. The Hall set have current flowing *via* diagonally-opposite, *i.e.*, non-adjacent, contacts with the voltage measured on the perpendicular diagonal pair. The latter are mostly used at finite magnetic field to obtain the electron density but can also give rough estimates for anisotropy under the right conditions (see Supplementary Figures S3 & S6). Here, we focus on the former to investigate the anisotropy. The van der Pauw (vdP) resistance is defined as $R_{hi,jk} = V_{jk}/I_{hi}$ where h and i denote adjacent current probes and k and j adjacent voltage probes. For example, $R_{AB,DC}$ is measured by passing current between A and B and probing the resulting voltage between D and C. The 16 permutations in the van der Pauw set fall into four groups of four permutations where the four are all equal, *i.e.*, $|R_{AB,DC}| = |R_{BA,DC}| = |R_{AB,CD}| = |R_{BA,CD}|$ because we use a.c. lock-in techniques to drive current and measure the resulting voltage. From here forth we focus entirely on the four physically distinct probe allocations in terms of measurement topology in the van der Pauw set: $R_{AB,DC}$, $R_{BC,AD}$, $R_{CD,BA}$ and $R_{DA,CB}$.

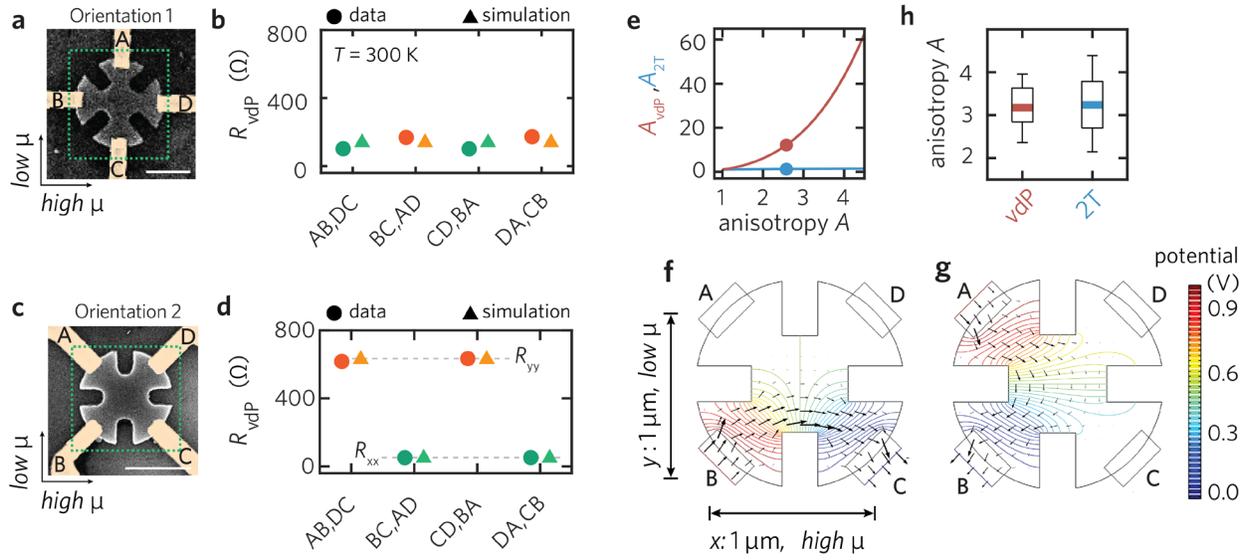

**Figure 3. a/c** SEM images of two van der Pauw (vdP) cloverleaf devices where the etch pattern and contacts are rotated by 45º from **a** Orientation 1 to **b** Orientation 2. The four contacts are labelled A-D running counter-clockwise and the shape of the pristine nanofin prior to etching is indicated by the green dashed line. The scale bars in **a/c** are 1 μm. **b/d** van der Pauw resistance $R_{vdP}$ for all four measurement configurations with experimental data as circles and simulated values as triangles. Simulation parameters were **b** carrier density $n = 5.5 \times 10^{17}$ cm$^{-3}$ and **d** $n = 3.4 \times 10^{17}$ cm$^{-3}$, with mobility $\mu = 3000$ cm$^2$/Vs and anisotropy $A = 2.6$ for both **b** and **d**. **e** Simulated resistance anisotropies $A_{vdP}$ (brown) and $A_{2T}$ (blue) vs underlying mobility anisotropy $A$ for the device in **c**. The two data points (circles) correspond to the experimental data for this device. **f/g** Simulated current density (arrows) and equipotential lines (colormap) for a cloverleaf device for horizontal **f** and vertical **g** measurements in Orientation 2. **h** Mobility anisotropy $A$ for vdP and two-terminal (2T) measurements for nine separate devices in Orientation 2 (including **c**) obtained at $V_{BG} = 0$ at room temperature.

Figures 3a and c show two devices fabricated in Orientation 1 and Orientation 2 matching the structures discussed in Fig. 1a/b. If polytypism causes transport anisotropy in our nanofins we expect to observe a clear asymmetry in the van der Pauw resistance measurement in Orientation 2 and no asymmetry in Orientation 1. This is indeed what we observe in Fig. 3b/d. The circles in Fig. 3b indicate the four $R_{vdP}$ measurements obtained for Orientation 1, with $R_{vdP}$ differing by 70 Ω between the two measurement configurations for values of 100 Ω and 170 Ω. We attribute this small asymmetry to a slight offset in the alignment between the crystallographic directions and the etch mask – see Supporting Information for further discussion. A vanishing-anisotropy is expected as these measurements are all diagonal, *i.e.*, at 45°, to the growth axis/stacking fault planes, and thus geometrically equivalent from a transport perspective. Additionally, Fig. 3b demonstrates agreement with Schmid et al.,[20] namely that data consistent with isotropic transport can be obtained from van der Pauw measurements under certain orientations even if the underlying material is anisotropic. The $R_{vdP}$ data for Orientation 2 are shown in Fig. 3d and present a stark contrast to that for Orientation 1 – the $R_{vdP}$ for the AB,DC and CD,BA configurations are 12 times higher than for the BC,AD and DA,CB configurations. The results in Fig. 3d point to a significant resistance anisotropy related to polytypism since the two low resistance configurations have current flowing parallel or perpendicular to the stacking fault planes while the two high resistance configurations are *vice versa*. Consistent behavior was observed across all eleven devices characterized in this study.

Before we address the underpinning crystallographic aspects, we will perform some additional analysis for the devices in Fig. 3 to more rigorously define and quantify the observed anisotropy. There are two obvious ways to define anisotropy in the measurements above. The first is an experimentally measured van der Pauw resistance anisotropy $A_{vdP} = R_{yy}/R_{xx} \geq 1$, for which we define the $x$ and $y$ directions and assign the contacts A-D in Fig. 3a/c such that $R_{xx} = (R_{BC,AD} + R_{DA,CB})/2$ and $R_{yy} = (R_{AB,DC} + R_{CD,BA})/2$ and $R_{xx} < R_{yy}$. The second is an underlying mobility anisotropy $A = \mu_x/\mu_y$ where $\mu_x$ and $\mu_y$ are mobilities along the $x$ and $y$ directions, with $\mu_x \geq \mu_y$. The relationships between $R_{xx}$, $R_{yy}$ and $\mu_x$, $\mu_y$, and thus $A_{vdP}$ and $A$, are mediated by an anisotropic conductivity tensor under the Drude approximation:[55]

$$\boldsymbol{\sigma} = \frac{ne}{1+\mu_x\mu_y B^2}\begin{pmatrix} \mu_x & -\mu_x\mu_y B \\ \mu_x\mu_y B & \mu_y \end{pmatrix} \quad (1)$$

where $n$ is carrier density, $e$ is the electron charge and $B$ is the perpendicular magnetic field. The relationship between $A_{vdP}$ and $A$ has previously been calculated analytically for rectangular structures with infinitely small contacts at the corners.[55] However, these conditions are not fulfilled in most real nanoscale devices and certainly not in our case. As a result, we instead need to use a numerical model to compute the expected $A_{vdP}$ for a given $A$ for our device geometries, which we perform using finite-element modelling of our structures in COMSOL Multiphysics. The modelled device geometries are extracted from SEM images and approximated as shown in Fig. 3f/g. We define a 'general' mobility $\mu$ for the model, where $\mu^2 = \mu_x\mu_y$ following Bierwagen et al.[55] We chose $\mu = 3000$ cm$^2$/Vs and $n = 2 \times 10^{17}$ cm$^{-3}$ as default values based on our earlier work on InAs nanofins.[26] These values were used for all simulations unless otherwise stated. Note that $A_{vdP}$ as a function of $A$ does not depend on the absolute values of $\mu$ and $n$, only the relative values of $\mu_x$ and $\mu_y$ and the sample geometry. Figure 3e shows the modelled relationship between $A_{vdP}$ and $A$ for the device shown in Fig. 3c. This relationship is obtained by setting $A$, which determines $\mu_x$ and $\mu_y$, and then modelling $R_{xx}$ and $R_{yy}$ using the conductivity tensor from Eqn 1 to obtain $A_{vdP}$ (full details of the finite-element model are given in the Methods). We find that $A_{vdP}$ is extremely sensitive to changes in $A$ for the cloverleaf structure, with $A_{vdP} \approx 7$ for $A = 2$ and $A_{vdP} \approx 40$ for $A = 4$. The data in Fig. 3d corresponds to $A_{vdP} = 12$, which gives $A \approx 2.6$, i.e., $\mu_x \approx 2.6\mu_y$. This is indicated by the brown dot in Fig. 3e. Note that the simulation for Orientation 1 (triangles in Fig. 3b) shows no variation in $R_{vdP}$ even though the same $A = 2.6$ was used, highlighting that the small residual anisotropy in this instance is an experimental artefact.

The approximately exponential dependence of $A_{vdP}$ on $A$ for Orientation 2 underpins our earlier assertion for using van der Pauw measurements as an 'anisotropy amplifier'. An understanding of why this occurs is evident in Fig. 3f/g, where we plot the equipotential lines and current density from the finite element model for $A = 2.6$ in the measurement configurations that correspond to $R_{xx}$ and $R_{yy}$ measurement configurations, respectively. For the $R_{xx}$ configuration, the current preferentially flows in the higher mobility $x$ direction, confining the current to a region close to the current contacts B and C. The resulting voltage drop probed by A and D is small. In contrast, the current flows along the lower mobility $y$ direction in the $R_{yy}$ configuration, but crucially, the current distribution is much broader, i.e., a wider flow at lower density. This pushes the equipotential lines further outwards in the $x$ direction, resulting in a significantly higher voltage between D and C. It is this anisotropy-driven redistribution in current density and electrostatic potential that ultimately makes vdP measurements an effective tool for characterizing anisotropic transport.

The anisotropy amplifying effect of the vdP measurements becomes particularly evident when comparing to conventional two-terminal (2T) resistance measurements $R_{2T}^{h,i}$ where h and i indicate the two contacts used. We can calculate a two-terminal anisotropy $A_{2T} = R_{2T}^{AB}/R_{2T}^{BC}$ using the two-terminal resistances $R_{2T}^{BC}$ and $R_{2T}^{AB}$ from our finite element model as a function of $A$, giving the blue trace in Fig. 3e. In contrast to $A_{vdP}$, $A_{2T}$ is relatively insensitive to $A$. For example, at the $A = 2.6$ consistent with our device, we get $A_{2T} = 1.25$, an order of magnitude smaller than $A_{vdP}$. To provide a solid statistical footing for our result, we measured eleven separate devices matching Fig. 3c in shape and orientation, extracting the mobility anisotropy $A$ based on measurements of $A_{vdP}$ and $A_{2T}$, with the results presented in Fig. 3h. In the two-terminal resistance data a 700 Ω contact resistance, based on separate four-probe measurements, was assumed and corrected for. The $A_{2T}$ and $A_{vdP}$ measurements are in good agreement and the variance for $A_{vdP}$ is significantly smaller. We attribute this to the higher sensitivity of $A_{vdP}$ to $A$ (Fig. 3e) and the fact that the two-terminal data is more susceptible to fluctuations arising from, *e.g.*, device-to-device contact resistance variations. The anisotropy also affects the Hall measurements, as we discuss in the Supporting Information (Fig. S3/S5). In short, anisotropy produces an offset in the Hall voltage, *i.e.*, finite Hall voltage at zero magnetic field. However, for the magnitude of anisotropy we observe here, the Hall slope is only weakly affected and still gives accurate carrier density measurements.

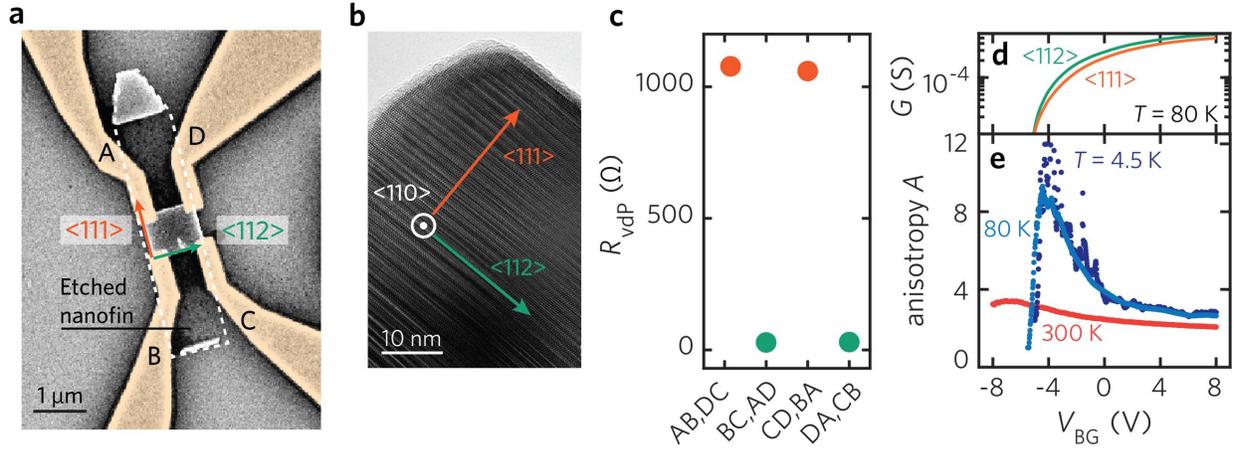

**Figure 4. a** False-colored SEM image of a rectangular van der Pauw structure (1 μm × 0.85 μm with 0.55 μm contact separation for all adjacent contact pairs) etched into a high aspect ratio nanofin. **b** TEM image of a nanofin from the same growth along the ⟨110⟩ zone axis revealing a high density of polytype stacking faults. **c** The van der Pauw resistance $R_{vdP}$ vs contact configuration shows a strong anisotropy with higher resistance when the current flows along the ⟨111⟩ direction. **d** Two-terminal conductance $G$ and **e** mobility anisotropy $A$ vs back-gate voltage $V_{BG}$. Data in **d** is obtained along the ⟨111⟩ (orange) and ⟨112⟩ (green) directions at $T = 80$ K. Data in **e** is obtained at $T = 4.5$, 80 and 300 K.

**Crystallographic aspects of the anisotropy:** The difficulty with the square nanofins is that we cannot track the growth axis orientation during transfer to the device substrate and we cannot easily identify the axis once the nanofin is in place either. For this reason, we now turn to our 5 μm × 1 μm nanofins, which enable growth axis tracking despite being too small to make good cloverleaf structures from. Figure 4a shows the device used to connect the transport anisotropy with crystallographic direction. The original full nanofin shape is indicated with the white dashed line. The high aspect ratio enables us to unambigiously determine the ⟨111⟩ and ⟨112⟩ crystallographic axes, which we can link to the directions perpendicular to and in the plane of the stacking faults, respectively, by transmission electron microscopy (Fig. 4b). We use an EBL-defined etch to obtain a small rectangular van der Pauw structure (1 μm × 0.85 μm) with four contacts at the corners as shown in Fig. 4a. We plot the measured vdP resistances for the four contact configurations in Fig. 4c with data representing transport along ⟨111⟩ and ⟨112⟩ shown in orange and green, respectively. We find approximately 40 times higher $R_{vdP}$ along the ⟨111⟩ direction corresponding to a lower mobility. Modelling the device geometry in Fig. 4a, the measured data corresponds to $A \approx 3$, which is consistent with the anisotropy we observe in our cloverleaf devices (Fig. 3h). Additional data from similar devices to that in Fig. 4a confirms this result and is shown in the Supporting Information.

Ultimately, our data in Fig. 4c enables us to tie the anisotropy to a higher mobility for transport in the plane of the stacking faults, *i.e.*, along ⟨112⟩, and a lower mobility for transport perpendicular to the stacking fault planes, *i.e.*, along ⟨111⟩. The increase in resistivity arising from transport across stacking fault planes is consistent with expectations from studies of longitudinal transport in InAs nanowires with pure WZ, pure ZB and WZ/ZB phase mixing.[43,44] The conductivity in bulk InAs (ZB) is isotropic. There are reports of anisotropy in III-V heterostructures such as modulation-doped (311) GaAs/AlGaAs,[62–64] $In_xGa_{1-x}As/In_xAl_{1-x}As$[65–68] and InAs/AlSb.[69] However, anisotropy in these systems is attributed to characteristics specific to the heterostructure such as non-isotropic strain,[65,70] anisotropic dislocations,[69] potential fluctuations due to inhomogeneous doping/elemental distributions,[66] or surface scattering[67] by periodic corrugations. For our nanofins, there are two possible contributions to the mobility that can cause anisotropic transport: the effective mass and the scattering time. The effective mass for ZB InAs is isotropic[71,72] and bandstructure calculations for WZ InAs[73] predict an effective mass ratio $m^*_{⟨112⟩}/m^*_{⟨111⟩}$ of approximately 1.5. This effect is small, it also runs counter to the reduced mobility along ⟨111⟩ suggesting that the observed anisotropy is not caused by effective mass anisotropy. Instead, we attribute the mobility anisotropy to a difference in scattering time caused by the polytypism. The conduction band edge for WZ sits 80-130 meV higher in energy than for ZB.[74,75] The WZ phase segments thus present a series of potential barriers along ⟨111⟩ with well-known ability to substantially increase the resistivity.[43,44] This interpretation is supported by additional measurements in Fig. 4d/e. Firstly, in Fig. 4d we plot the two-terminal conductance $G$ versus $V_{BG}$ for transport along the ⟨111⟩ and ⟨112⟩ axes at $T = 80$ K. The gap between these two traces is caused by the anisotropy $A$. The corresponding $A$ extracted in the vdP configuration is plotted against $V_{BG}$ as the light blue trace in Fig. 4e. As $V_{BG}$ is made more negative, depleting the nanofin, $A$ increases sharply reaching a peak at $V_{BG} = -4$ V. Thereafter an even sharper fall in $A$ occurs as $V_{BG}$ is made more negative in Fig. 4e, which arises as both traces in Fig. 4f converge towards $G = 0$ when the device reaches full depletion. We also plot $A$ versus $V_{BG}$ at higher (300 K) and lower (4.5 K) temperatures in Fig. 4e. The 300 K trace is much flatter with peak $A$ at $V_{BG} = -6$ V, indicative of thermal broadening and a temperature-induced threshold shift. The 4.5 K data has a higher peak $A$ at slightly more positive $V_{BG}$ for similar reasons; the additional fine structure arises from quantum interference effects. Similar behavior was observed for additional devices, as shown in the Supporting Information.

From an energetic perspective, our interpretation is that $V_{BG}$ modulates the Fermi energy $E_F$ relative to the conduction band edge. Decreasing $V_{BG}$ increases the separation between $E_F$ and the WZ segment band edge, increasing the effective barrier height along ⟨111⟩ and reducing the mobility in that direction. The opposite happens when making $V_{BG}$ more positive. The effect is exacerbated at lower temperatures due to reduced phonon scattering and because the sharper Fermi-edge increases the energy filtering effect of the WZ barriers.[75,76] Ultimately, we are able to tune $A$ by a factor of approximately 4 over the full back-gate range of 16 V. We estimate that we move $E_F$ by about 70 meV over this gate range, which is comparable to the band offset in InAs nanowires[75] (full details in the Supporting Information). We speculate that elevating $E_F$ well above the WZ barriers (highly positive $V_{BG}$) may further reduce the anisotropy. Note that a model of square-shaped WZ potential barriers in a pristine ZB lattice may be an oversimplification. A firmer quantitative analysis would require a more refined model that takes into account miniband formation within the ZB regions as well as thermionic and tunnelling transport contributions. Gaining better control over crystal phase in our nanofins would also be highly desirable as it would enable devices with, *e.g.*, single polytype barriers of controlled length for detailed studies of the impact of polarization charges on transport.[75,77]

**CONCLUSIONS**

We have demonstrated the post-growth shaping of free-standing InAs nanofins,[26] grown by selective-area epitaxy and transferred to sit flat on a device substrate, using a citric acid based wet-etch with a patterning resolution as low as 150 nm. The etch provides a route to making more complex multi-terminal devices, *e.g.*, van der Pauw cloverleaf structures, as well as enabling partial thinning of the nanofin to significantly improve and/or tailor the gate response. We exploit the high sensitivity to transport anisotropy of our cloverleaf structures to address the fundamental question of whether there is a measurable transport anisotropy arising from wurtzite/zincblende polytypism in these 2D III-V nanostructures. We obtain measured resistance anisotropies of order 12, which correspond to an underlying mobility anisotropy of order 2-4 at room temperature with no back-gate voltage applied. We find the higher mobility along the stacking fault planes (along $\langle 112 \rangle$) and the lower mobility perpendicular to the stacking fault planes (along $\langle 111 \rangle$). We show that the anisotropy varies based on temperature and carrier density, as controlled by the back-gate voltage. Our results highlight an important materials consideration that needs to be factored into the design of devices using free-standing 2D nanostructures grown by selective-area epitaxy.

**METHODS/EXPERIMENTAL**

**Nanofin epitaxy:** A detailed description of the template fabrication and nanofin epitaxy is provided in Ref. 26.

**Etchant:** We prepared the citric acid stock solution as follows. We mixed anhydrous citric acid powder (Chem-Supply) with deionized water at a mass ratio of 1:1 until it formed a homogeneous solution. The solution was stored in a sealed vessel.

The stock etch solution was prepared by mixing 250 μL of the citric acid stock solution above with 5 mL of 30% hydrogen peroxide solution (Honeywell). For the 1:2 and 1:5 dilutions we added a further 10 mL and 25 mL of deionized water. The etchant was then transferred to glass beaker sitting in a water bath at a constant temperature $T = 25°C$. The etchant was left to adjust to the water bath temperature for 120 s before commencing the etch. The sample chip was submerged and agitated in the solution during the etch process.

**Device Fabrication:** The device substrates were 300 μm thick, 2 inch (100) Si wafers (SVMI) doped *n*-type to 0.001−0.005 Ωcm. They featured a 130 nm thick thermally grown $SiO_2$ layer on the front-side, which acted as a gate insulator for the doped Si wafer (back-gate). Ti/Au bond-pads, interconnects and alignment markers were fabricated on the front side of the wafer by one round of photolithography (bond-pads, interconnects) and electron-beam lithography (alignment markers). These wafers were then cleaved into small 3.5 mm × 5.5 mm device chips that featured 24 device fields with four contact at the corners. These chips were sonicated in a mixture of acetone and 2-propanol prior to use. Nanofins were then transferred from the growth substrate with a custom-built micromanipulator system featuring an ultra-fine needle (100 nm tip radius, American Probes) mounted on a robot arm (Eppendorf) and a motorized stage (Zaber) under an optical microscope (Leica).

For the etch masks we spun PMMA A5 (Microchem) at 5000 rpm for 60 s, which we exposed by EBL (20 kV, 20 μm aperture) using a Raith-150 Two electron-beam lithography system to a dose of ~300 μC/cm$^2$ after baking the resist at 180°C for 5 min. After development in a 1:3 mixture of methylisobutylketone and 2-propanol for 60 s, the exposed sections were cleaned with a 30 s $O_2$ plasma ashing step (340 mTorr, 50 W). The citric acid etch step then followed, as described in the main text.

Metal contacts were defined by an EBL process equivalent to that given above. The InAs nanofins were passivated by a sulfur-based passivation solution prior to metal-evaporation. The passivation solution was prepared by adding 2.4 g of sulfur to 25 mL of 20% aqueous ammonium sulfide solution to create a stock solution. This stock solution was diluted at 1:100 with deionised water for the final passivation solution. After passivating the devices for 120 s at 40°C, the sample was immediately loaded into a thermal vacuum evaporator. Typical contact metal stacks were 5/135 nm Ni/Au for the 60-70 nm thick nanofins (Figs. 3/4) and 6/194 nm Ni/Au for the 150 nm thick nanofin in Fig. 2. The final devices were obtained by lift-off in acetone at 60°C and rinsing in 2-propanol.

**Electrical measurements:** All electrical measurements were carried out under vacuum in an Oxford Instruments Heliox VT $^3$He cryostat to ensure consistency between room temperature and low-temperature measurements. The cryostat featured a superconducting magnet, enabling Hall measurements in a perpendicular magnetic field of up to 2 T. We used standard lock-in measurements at a frequency of 77 Hz using a Stanford Research Systems SRS830 lock-in amplifier. A Keithley 2401 source measure unit supplied the dc back-gate voltage $V_{BG}$.

The data in Fig. 2f was obtained by sourcing 3 V (rms) ac into a 1:10000 voltage divider, giving a $V_{SD}$ of about 300 µV. The currents $I_P$, $I_E$ and the corresponding voltage drops were recorded while sweeping $V_{BG}$ to obtain resistances for the two sections. For the van der Pauw measurements as in Figs. 3b/d and Fig. 4c, we sourced a variable $V_{SD}$ (0 to 300 µV) to one of the two-terminal contacts and measured the resulting current $I_{SD}$ to ground at the other. Simultaneously we measured the voltage drop $V_{vdP}$ across the opposite vdP contact pair. This was repeated for all four contact configurations. The two terminal and vdP resistances were then determined from linear fits to $I_{SD}$ *versus* $V_{SD}$ and $I_{SD}$ *versus* $V_{vdP}$ data, respectively. The measurements shown in Fig. 4d/e were performed by sourcing $V_{SD}$ = 300 µV into one two-terminal contact while measuring the current to ground through the other. We simultaneously monitored the vdP voltage across the opposite contact pair, giving the two-terminal and vdP resistances.

**Finite element simulations** The van der Pauw devices were simulated with a two-dimensional COMSOL Multiphysics 5.1 model using the *electric currents* (ec) module. We used a stationary study with a relative tolerance of $10^{-5}$. The Au contacts were modelled with an isotropic conductivity of $8.5 \times 10^8$ S/m while the InAs channel was assigned the conductivity tensor described in the text. We used a thickness parameter of 60 nm matching the typical nanofin thickness for our cloverleaf devices. The outside surface of each two-terminal contact was a COMSOL terminal of voltage type that was kept at constant voltage at the source (1 V in Figs. 3f/g) and 0 V at the drain. The van der Pauw contacts were COMSOL floating potentials that allowed the extraction of their potential from the simulation.

**AUTHOR INFORMATION**

**Corresponding Author**

* adam.micolich@nanoelectronics.physics.unsw.edu.au

**Current Addresses**

§ Microsoft Quantum Lab Delft, Delft University of Technology, 2600 GA Delft, The Netherlands.

† Hunan Key Laboratory of Nanophotonics and Devices, School of Physics and Electronics, Central South University, 932 South Lushan Road, Changsha, Hunan 410083, P. R. China.

## ACKNOWLEDGMENT

We thank R.W. Lyttleton for helpful discussions on the etch chemistry, and N. Shahid and S. Naureen for contributions to the development of the selective-area epitaxy process used for the nanofin growth in this work. This work was funded by the Australian Research Council (ARC) and the University of New South Wales. This work was performed in part using the NSW and ACT nodes of the Australian National Fabrication Facility (ANFF) and the Electron Microscope Unit (EMU) within the Mark Wainwright Analytical Centre (MWAC) at UNSW Sydney.

## ASSOCIATED CONTENT

**Supporting Information Available:**

(S1) Additional AFM data for thinned down devices and electrical data from Fig. 2f on a logarithmic scale. (S2) Additional data on isotropy of the etch. (S3) Additional resistance data in the Hall (diagonal) configuration as well as Hall voltage data in the two orientations. (S4) Simulation data on angle-dependence of van der Pauw measurements. (S5) Simulation data on angle-dependence of Hall measurements. (S6) Simulation data on extracting transport anisotropy from two-terminal diagonal measurements (Hall orientation). (S7) Additional electrical data in support of Figure 4. van der Pauw resistances for an additional square-shaped device, $T$-dependent anisotropy $A$ for an additional cloverleaf device, $A_{vdP}$ and $A$ at 300 K for additional devices and $E_F$ vs $V_{BG}$ derived from Hall studies.

This material is available free of charge *via* the Internet at http://pubs.acs.org.